# Enhancing Engagement and Learning in Computing Education: Automated Moodle-Based Problem-Solving Assessments

Charith Jayasekara[1]; Carlo Kopp[1]; Vincent Lee[2]; Chetan Arora[1];

[1]Department of Software Systems and Cybersecurity, Faculty of Information Technology, Monash University, Australia

[2]Department of Data Science and AI, Faculty of Information Technology, Monash University, Australia

**Abstract**

*This paper presents the design and refinement of automated Moodle-based Problem-Solving Assessments (PSAs) deployed across large-scale computing units. Developed to replace traditional exams, PSAs assess applied problem-solving skills through parameterised, real-world tasks delivered via Moodle's quiz engine. Integrated with interactive workshops, this approach supports authentic learning, mitigates academic integrity risks, and reduces inconsistencies in marking. Iterative improvements have enhanced scalability, fairness, and alignment with learning outcomes. The model offers a practical and sustainable alternative for modern computing and engineering education.*

***Keywords:*** *problem-solving, computing education, Moodle assessments, academic integrity, scalable assessment*

## 1. Introduction
### 1.1. Background and Motivation

Tertiary-level computing units, particularly those in systems and networks, have historically followed a conventional assessment structure, combining mid-semester tests, major programming assignments, and a high-stakes final exam. While this model once suited smaller cohorts and traditional delivery modes, it is increasingly strained by growing enrolments, digital delivery formats, and rising expectations for industry-aligned, applied learning (Villarroel et al., 2020). Core and key fundamental units such as Operating Systems, Computer Architecture and Networks demand both theoretical understanding and practical problem-solving. However, given the diversity in academic pathways and varying levels of STEM preparation among incoming students, it has become essential to provide additional scaffolding and design adjustments to support equitable learning outcomes (Moghaddam et al., 2021). Outdated textbook-style assessments and disconnected content further limit students' ability to develop the reasoning required for contemporary computing challenges (Villarroel et al., 2020). This situation highlights a more fundamental concern: Are our assessments still measuring capability in an authentic way, or are they simply rewarding recall and exam technique? Increasingly, educators are called upon to design evaluation models that reflect cognitive demands and simulate real-world problem-solving. The shift outlined in this paper was driven by the need to embed assessments that not only evaluate learning but actively stimulate it, activating higher-order thinking, conceptual integration, and realistic computing contexts (Ben Ghalia, 2024).

1.2. Rationale for PSA Model Adoption

Traditional final exams often promote surface learning through last-minute cramming and memorisation, with limited engagement throughout the semester (French et al., 2024). Their rigid, time-pressured formats constrain assessment of reasoning and problem-solving, and disadvantage students who struggle under high-stakes conditions. At the same time, non-invigilated online assessments face rising concerns about academic integrity, particularly with the emergence of generative AI tools, while large-scale paper-based exams are becoming increasingly unsustainable (Bittle & El-Gayar, 2025). To address these challenges, Moodle-based Problem-Solving Assessments (PSAs) were introduced as a scalable, authentic alternative. Rather than digitising existing exams, the model was designed from the ground up to support ongoing learning through parameterised, scenario-based tasks that evaluate students' ability to apply knowledge in meaningful contexts. Delivered via Moodle's quiz platform, PSAs incorporate conditional logic, automated marking, and real-world framing to enhance engagement, reduce misconduct, and minimise marking overhead (Gamage et al., 2019). Each PSA is embedded within a problem-solving workshop, creating a cohesive learning and assessment environment that promotes continuous participation. The model aims to enhance authenticity, academic integrity, and cognitive depth while shifting the focus from rote recall to real-world application. This paper outlines how the PSA framework was iteratively refined and integrated across multiple computing units to overcome the limitations of legacy assessments.

2. **Assessment Design and Implementation**
    2.1 Model and Delivery Philosophy

The Moodle-based PSA model was purposefully developed to tackle challenges in authenticity, cognitive rigour, academic integrity, GenAI resilience, and scalability in computing assessments. Unlike conventional digital exams, PSAs focus on deeper learning by engaging students in parameterised, scenario-based tasks aligned with real-world computing contexts. Grounded in principles such as the spacing and testing effects, PSAs are scheduled at key points during the semester (Weeks 5, 9, and 12). Each assessment reinforces conceptual understanding, scaffolds complexity, and helps build problem-solving fluency through spaced retrieval and progressive challenge (Sisti et al., 2007).

Parameterisation is one of the key design pillars here. Students receive equivalent but unique versions of each task with varied inputs, values, or sequences, ensuring fairness while maintaining challenge and academic integrity. All PSAs are delivered within weekly workshops. Sessions are supervised on campus or via Zoom (with focus mode), with mandatory full-screen sharing and camera use. Randomised parameters and shuffled orderings eliminate the need for traditional exam halls or seating plans. This effectively mitigates the risks of collusion or answer sharing. Workshops also serve as practice environments, where students solve problems similar to those in PSAs, strengthening alignment between learning and assessment. This format enhances participation and engagement, resulting in increased workshop attendance throughout the semester. Each PSA is designed to align directly with unit-level learning outcomes.

Questions are structured to assess threshold concepts in realistic scenarios, supporting both summative evaluation and formative development. The effectiveness of this alignment is further analysed in Section 4, along with student feedback and performance data.

2.2 Question Types and Automation

The PSA model leverages a diverse set of Moodle question types to simulate realistic computing tasks while supporting automated marking at scale. Each question is developed to assess applied knowledge, critical thinking, and procedural problem-solving, rather than surface-level recall. The following question types are prominently used in PSAs.

- Numerical Calculated (short and long)
- Multiple-Choice Single-Answer (calculated and non-calculated)
- Multiple-Choice Multiple-Answer
- Cloze (Embedded Answers)
- Match and Select-Type

In our PSA implementation, numerical and calculated numerical questions were used for algorithmic and system-level tasks, i.e. memory calculations, disk access time estimation, and scheduling metrics. Each student received different parameter values, ensuring uniqueness and reducing the risk of collusion. Multiple-choice (mostly calculated) single-answer questions were employed for lower-difficulty checkpoints, offering quick validation of discrete concepts like process states or I/O types. Multiple-choice multiple-answer questions were intentionally designed to be more cognitively demanding. These assessed layered conceptual understanding by requiring all correct responses and penalising incorrect selections with zero marks, removing the possibility of guessing. Cloze (embedded) questions were central to our scaffolding approach, presenting multi-step tasks in a single narrative such as breaking down paging operations or multi-stage computations, while maintaining continuity and logical flow. Match and select-type formats were used for classification and mapping exercises, such as linking system components to their functions or command types to expected outputs. Calculated multichoice questions blended computation with interpretative judgment such as choosing the optimal system decision following a numerical calculation, allowing us to probe students' applied reasoning. Together, these question types form a multi-layered, problem-rich assessment environment spanning Bloom's taxonomy. Their comparative strengths and student responses are further discussed in Sections 4 and 5.

3. **Iterative Refinement Process**

Over successive offerings, Moodle-based PSAs have undergone deliberate refinements in structure, content, and cognitive complexity. These enhancements were based on internal teaching team reviews and performance outcomes. The goal was to increase the diagnostic value and fairness of the assessments while maintaining strong alignment with weekly workshops and unit learning outcomes.

Table 1 - Summary of iterative changes made over the offerings.

| Change Implemented | Remarks |
|---|---|
| Increased number of questions | Better matched workload expectations and allowed wider topic coverage. |
| Expanded scenario complexity | Moved from straightforward questions to layered, multi-concept scenarios. |
| Greater use of parameterised questions | Created personalised variants while preserving consistent cognitive load. |
| Multi-answer MCQs with strict marking | Required confident selection of all correct answers only; reduced guessing. |
| Widespread use of Cloze formats | Integrated multi-step reasoning within a single task to assess problem-solving processes. Enabled controlled build-up of complexity, supporting both accessibility and challenge. |
| Structural consistency with workshop content | Maintained consistent use of visuals, formats, and terminology used in teaching materials. |

These changes were applied across all units using the PSA model, not tied to a specific subject or level. The increase in the number of questions was carefully balanced against expected completion times. Complexity was introduced by embedding conditional constraints, diagram-based analysis, and computational tasks within problem contexts. Multi-answer MCQs and Cloze questions were particularly valuable in assessing critical thinking, as they required both recall and application in a structured way. Each enhancement was designed to maintain assessment rigour while promoting fairness, authenticity, and alignment with learning goals.

4. PSA vs Legacy Assessments

The shift from centralised, high-stakes exams and assessments to integrated Moodle-based PSAs offers a fundamentally different approach to evaluating student learning. The tabel 2 contrasts the PSA model with traditional final exams across key dimensions.

Table 2 - Comparison of Traditional Exams vs. PSA Model

| Aspect | Traditional Final Exams | Moodle-Based PSAs |
|---|---|---|
| Delivery and Scheduling | Centrally scheduled, time-limited exam | Embedded in scheduled workshops across semester (Weeks 5, 9, 12) |
| Assessment Format | Static, same set of questions for all students | Parameterised questions with randomised values and shuffled options |

| Cognitive Load and Spacing | High extraneous load, one-off assessment | Lower load via familiar setting, multiple spaced assessments supporting retention |
|---|---|---|
| Skill Focus (Bloom's Taxonomy) | Recall and basic interpretation (Remember, Understand) | Application, analysis, and evaluation via scenario-based and layered problem solving |
| Question Types and Scaffolding | Mostly short answers and calculations | Includes Cloze (embedded), multi-response MCQs, match/select, and calculated types |
| Academic Integrity | Requires external invigilation, vulnerable in remote settings | On-campus supervision or Zoom-based supervision, screen sharing, and per-student variation reduce misconduct |
| Marking and Feedback | Manual marking prone to delays and variation | Auto-marked via Moodle, consistent partial marking logic for nuanced evaluation |
| Student Engagement and Experience | Engagement peaks near exam time | Sustained engagement throughout semester, embedded in active workshop learning |
| Operational and Logistical Load | High administrative overhead for exams | Minimal logistics; uses existing teaching infrastructure |

Traditional assessments, whether time-limited exams, take-home assignments, or weekly quizzes, tend to emphasise memory recall, create uneven learning rhythms, and struggle with growing concerns around academic integrity and generative AI misuse. PSAs, in contrast, are designed to promote sustained engagement, assess higher-order thinking, and ensure authenticity in student responses. Embedded directly within scheduled workshops, PSAs replace the pressure of a one-time exam with spaced, supervised assessment opportunities that align with weekly teaching rhythms. This approach promotes long-term memory retention through iterative reinforcement and allows students to demonstrate conceptual understanding in familiar environments, mitigating the cognitive overload and anxiety often associated with traditional exams. From a cognitive perspective, PSAs rely on carefully scaffolded formats such as Cloze (embedded logic), calculated numericals, and multi-response MCQs. These are structured around real-world problem-solving scenarios, moving beyond the recall-oriented nature of both final exams and traditional quizzes. While quizzes are often used as low-stakes formative checks or basic recall tools, PSAs are high-stakes, summative assessments designed to probe applied knowledge, reasoning, and abstraction. Unlike quizzes, PSAs incorporate parameterised inputs, adaptive sequencing, and summative integration with teaching, offering deeper insight into student capabilities and performance across the semester.

Traditional take-home assessments, while convenient, have also become increasingly susceptible to unauthorised assistance and AI-generated responses. The PSA model counters these risks through a combination of real-time invigilation and structural variation, ensuring students must reason through tasks independently. Importantly, this is achieved without relying on external proctoring tools, making the model scalable across both online and on-campus cohorts. Marking reliability is enhanced through automation within Moodle, using partial scoring rules and calculated item logic to deliver timely, consistent feedback. This reduces long term administrative overhead and supports grading transparency, unlike legacy methods that are often subject to delays, inconsistencies, or limited feedback loops. Ultimately, PSAs represent a sustainable alternative to traditional assessments by tightly integrating learning and evaluation. They use authentic, dynamic tasks to reinforce conceptual mastery, promote academic integrity, and reduce logistical complexity. The following table (Table 3) highlights key differences between legacy assessment types and PSAs, including observed variations in cognitive load and task difficulty.

Table 3 - Comparison of final exam questions and PSA questions

| Exam Question | PSA Question | Remarks |
| --- | --- | --- |
| Explain Software-Defined Networking (SDN) | <ul><li>Control Plane A, hosted on a Nexus-X1 controller, handles basic access control rules. It is capable of processing up to 1500 instructions per second, and each basic rule consumes 100 instructions.</li><li>Control Plane B, running on a Nexus-X2 controller, manages intermediate routing rules for departmental traffic, with a processing capacity of 2550 instructions per second. Each of these rules requires 150 instructions.</li><li>Control Plane C, the most advanced, operates on a Nexus-X3 controller and is responsible for handling complex application-aware policies used in telemetry, AI-driven routing decisions, and cloud-integrated firewalling. It can process 3000 instructions per second, and each complex rule consumes 200 instructions.</li></ul>While the SDN control planes manage the dynamic flow rule installation, a central Cisco ASR 1000 edge router is responsible for forwarding all real-time traffic. During peak business hours, this router handles an incoming traffic load of 22,900 packets per second. It has a high-performance service capacity of 25,000 packets per second under normal operating conditions.<br>a) Based on this setup, Calculate the maximum number of flow rules that can be processed collectively per second by the SDN control infrastructure.<br>b) Compute the router's traffic intensity, defined as the ratio of packet arrival rate to service rate, and round your final answer to three decimal places.<br>c) Based on the calculated traffic intensity, whether the router is approaching congestion risk or is operating within a safe load range. | This question significantly elevates cognitive demand by embedding multiple interrelated quantitative problems within a realistic enterprise networking scenario. It moves beyond rote recall by requiring multi-step calculations involving throughput, resource allocation, and performance evaluation. The complexity is high (Bloom's: Analyse, Evaluate) and appropriately reflects real-world SDN operations |
| What is the difference between paging and segmentation in the | A computing cluster utilised for data analysis shows the subsequent free memory partitions after a day's operation: | contextualised applied question |
| context of virtual memory systems? | 700K, 800K, 400K,600K, and 200K. During off-peak hours, jobs requiring 650K, 250K, and 750K of memory need to be scheduled.<br>If the scheduling system initially uses the<br>First-fit method<br>and then switches to the<br>Best-fit method<br>after accommodating the first job (because of an optimization algorithm), determine the<br>total unallocated memory<br>left in the system after scheduling all jobs (inKB). | assessing memory allocation strategies through numerical reasoning (Bloom's: Apply, Analyse) |
| Calculate the usable storage capacity in a RAID0 storage array, comprising eight 3 Terabyte disk drives?<br>Select one:<br>a. 4 TB<br>b. 8 TB<br>c. 24 TB<br>d. None of the above. | Calculate the cost per usable Terabyte of storage for a RAID1 storage array, comprised of 6 disk drives, each with a capacity of 4 Terabytes and a cost of $59. | Adds practical depth by requiring cost-per-TB calculation, shifting from recall to applied understanding of RAID systems (Bloom's: Apply). |

4.1 Final Exams and PSA Averages

Figure 1 - Result variation over exams and PSA for units

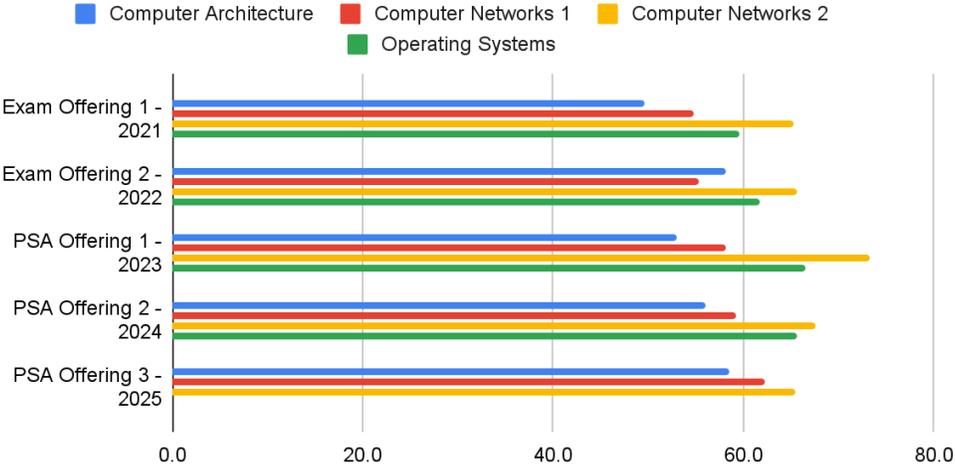

Despite incorporating more challenging question types and emphasising critical thinking, PSAs consistently yielded higher average scores than traditional final exams across all study areas examined in this study. As outlined in Section 4, PSA tasks were designed to assess deeper learning through scenario-based application, layered reasoning, and analytical decision-making. However, rather than leading to lower performance, this shift in cognitive demand appears to have supported stronger student outcomes. As illustrated in the figure 1, PSA averages surpassed final exam averages across most of the offerings, indicating that students not only adapted to the problem-solving model but performed more effectively when assessments were spaced, scaffolded, and embedded within familiar learning environments. These gains suggest improved alignment with learning outcomes, as well as greater retention and transfer of knowledge. This trend is reinforced by student feedback, which repeatedly emphasised the clarity, relevance, and fairness of PSAs. Comments in Student Evaluations of Teaching Units (SETU) highlighted the value of continuous assessment, supportive workshop-based delivery, and the opportunity to demonstrate understanding through practical, real-world scenarios, factors that likely contributed to the improved performance metrics.

## 5. Observations and Teaching Reflections

5.1 PSA results analysis

The following figure illustrates the variation in PSA averages across multiple offerings.

Figure 2 - average score variation of PSA 1, 2 and 3 for multiple offerings for Computer Networks 1 & 2, Computer Architecture and Operating Systems.

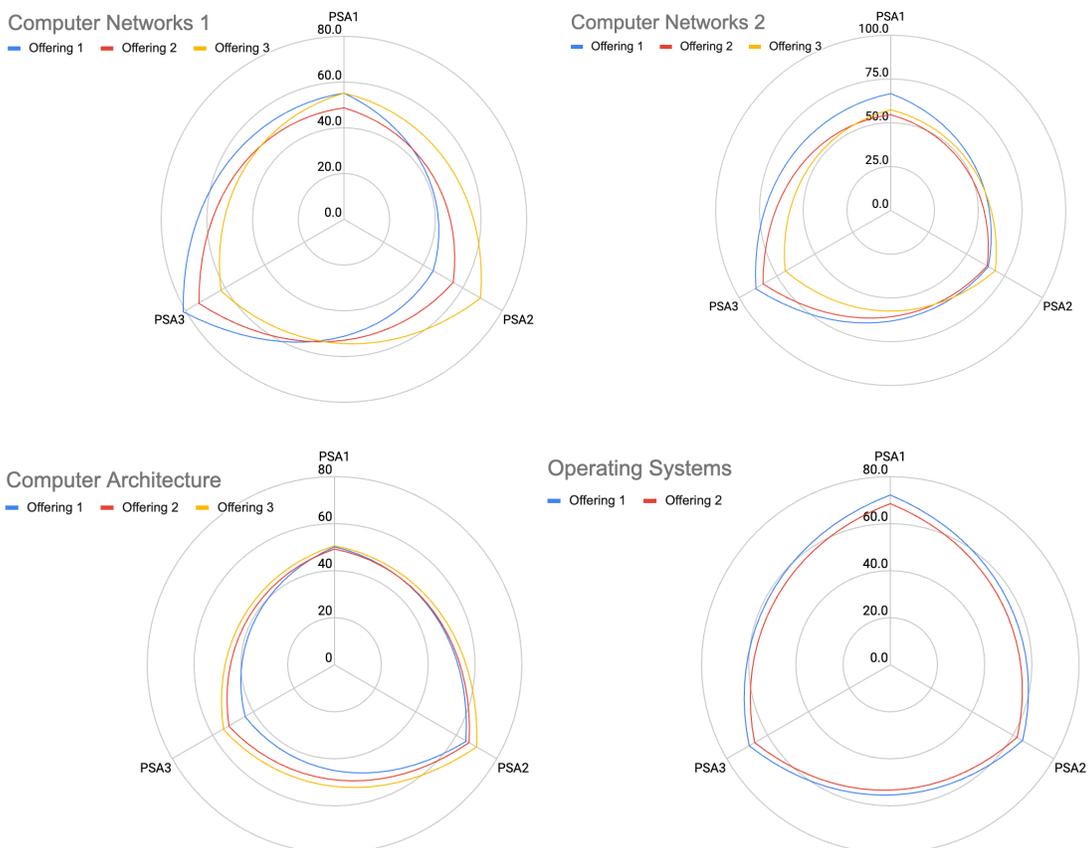

Performance trends across multiple offerings highlight the effectiveness of the PSA model in promoting learning, even as question complexity increased over time. Despite introducing layered Cloze formats, embedded conditions, and multi-step reasoning tasks, average scores generally remained stable or improved, indicating strong student adaptability and growing alignment between teaching and assessment.

Following is a key patterns observed across the units and offerings

- In Computer Architecture, PSA1 scores averaged around 49–50 but improved to 70.1 by the third offering, highlighting how early assessments helped students align their preparation and expectations.
- Across units, lower PSA1 performance prompted deeper engagement, with PSA2 scores showing broad improvement following feedback and workshop support.
- In Operating Systems, a high PSA1 average (72.4) was followed by a slight PSA2 dip, suggesting that early success may reduce perceived urgency to prepare further.
- PSA3 scores were generally stable or rising, with Computer Networks showing consistent improvement, reflecting the benefits of spaced assessment and format familiarity.
- Although complexity and question volume increased over time, average scores improved, indicating effective adaptation through refined pacing and alignment.
- Strong outcomes in later PSAs, despite higher-order and multi-step demands, confirm the value of integrated workshops and authentic, practice-based design.

Overall, the data demonstrate that students are capable of performing better in more authentic and demanding assessments when those are thoughtfully integrated into the teaching cycle. The PSA model not only preserves academic rigour but also fosters progressive improvement through repeated engagement and timely feedback.

5.2 Engagement Patterns and Workshop Dynamics

The integration of PSAs within workshop sessions has significantly enhanced student engagement. Rather than viewing workshops as secondary to lecture content, students now perceive them as essential preparation for the assessments. This has led to increased attendance, more focused peer interaction, and deeper engagement with the content. During PSA weeks, the intensity of focus observed across physical and virtual classrooms highlights how the assessment model has elevated the perceived value of formative sessions. Student feedback from SETU further confirms this shift, with recurring comments praising the relevance of PSAs and noting how well they align with the learning activities conducted throughout the semester. Students also reported reduced anxiety compared to traditional exams due to the distributed nature and integration of assessments with familiar settings. This is further supported by consistently high SETU scores for the item "The assessment in this unit allowed me to demonstrate the learning outcomes", with all units averaging above 4 out of 5 and a combined mean of 4.27.

Following are a few example feedback received from students

Table 4 - Student feedback related to PSAs and assessment structure

| Positive | Negative |
|---|---|
| "I really liked the layout of the assessments, as they would test my knowledge effectively whilst not being lumped together (for example, the PSA's all testing different parts of the unit)" | "The first PSA was unreasonably difficult. I think it should more closely resembled the second two PSAs, with questions more closely related to the workshops." |
| "Having 3 PSAs instead of a final exam is great as it reduces stress at the end of the semester and ensures I am keeping up to date on all the content." | "More revision questions. More clarity in PSA questions." |
| "evenly spreading the workload across 12 weeks was pretty effective when it comes to engaging students to learn," | "I don't like the assessments" |
| "Applied classes were nice, idea of PSAs over assignments/exam also good," | "PSAs are too tough." |
| "Workshop is very efficient when explaining sample question that might come in PSA" | "Maybe more practice question types for PSA" |

5.3 Platform and Infrastructure Support

The successful delivery of PSAs at scale has demonstrated the robustness of the Moodle platform when used strategically. Key features such as question banks, parameterised variables, timed delivery, and randomisation have supported both pedagogical goals and logistical execution. The ability to run assessments across multiple campuses and modes simultaneously has proven particularly valuable in hybrid learning contexts. Institutional support has been crucial, particularly in managing large-scale Zoom sessions, and ensuring stable Moodle / Zoom performance during peak times. Despite some limitations (e.g., lack of native SEB integration with Zoom), the overall infrastructure has supported a seamless and equitable assessment experience.

6. Discussion

The adoption of Moodle-based PSAs marks a shift from traditional assessment models towards more integrated and authentic evaluation. Rather than treating assessment as an isolated, summative event, the PSA model embeds it within the learning cycle, using scaffolded, scenario-driven tasks to measure not only what students know but how they apply that knowledge in realistic contexts. Grounded in pedagogical principles such as spaced retrieval, cognitive scaffolding, and contextualised reasoning, PSAs promote continuous engagement and deeper cognitive processing. While Moodle quizzes are often used for quick formative checks or surface-level MCQs, PSAs differ fundamentally in purpose and design. PSAs are not generic quizzes, they are structured, high-rigour assessments constructed from layered, parameterised question sets tied to weekly workshops. They incorporate numerical reasoning, embedded logical sequences (Cloze), multi-response MCQs, and scenario-based problem-solving aligned with learning outcomes. This design ensures higher-order skill evaluation and minimises rote learning or AI-enabled shortcuts. Importantly, the PSA model shows promise for scalability beyond computing. Disciplines requiring applied reasoning, data interpretation, and structured decision-making (e.g. engineering, analytics, health sciences) can adopt similar LMS-enabled designs. However, generalisability must be context-aware. While our study involved relatively technical units, the core design principles such as spaced engagement, scenario-based reasoning, parameterisation are agnostic to content and can be flexibly adapted. That said, not all disciplines or student groups may respond uniformly. Cultural attitudes toward assessment, prior exposure to active learning, and language proficiency may influence how students engage with PSAs. Therefore, effective deployment requires alignment with local teaching practices, student preparation, and support systems. As discussed in Section 5, ongoing teaching team collaboration and student feedback are critical for iterative refinement. Despite challenges such as system limitations or the time needed to develop high-quality banks, PSAs have demonstrated strong engagement, low misconduct rates, and improved alignment between learning and evaluation. This reinforces their viability as a future-facing assessment model when implemented with contextual sensitivity and pedagogical intent.

7. **Conclusion**

Moodle-based PSAs offer a viable, pedagogically sound alternative to traditional assessments, especially in large-scale higher education contexts. By replacing high-stakes, time-constrained exams with distributed, authentic evaluations embedded in teaching cycles, the model supports deeper learning, better engagement, and more robust academic integrity. Unlike standard LMS quizzes, PSAs involve complex, multi-step reasoning tasks uniquely generated per student. This elevates them beyond surface-level checks into assessments of applied capability and reasoning. As results across computing units indicate, students can succeed in cognitively demanding tasks when assessments are scaffolded, spaced, and well-aligned with learning activities. Although generalisability depends on local context and student backgrounds, the underlying approach of realistic, structured problem-solving supported by parameterisation and automation is broadly transferable. With thoughtful adaptation, PSAs can serve disciplines facing similar assessment challenges. Ultimately, the PSA model reframes assessment as part of the learning journey rather than a final checkpoint. As educational environments evolve amidst technological and pedagogical shifts, PSA-like designs provide a sustainable, integrity-conscious path forward, one that balances rigour, scale, and authentic learning.

8. **Acknowledgments**

We acknowledge the Traditional Custodians of the lands where this work took place, and pay our respects to their Elders past, present, and emerging. We thank all the teaching teams for their vital role in shaping the PSA model, and the students whose engagement and feedback both informal and via SETU guided its ongoing refinement. This study was approved by the Monash University Human Research Ethics Committee (Project ID: 47761).